\documentclass[aps,prd,twocolumn,superscriptaddress,amsmath,amssymb]{revtex4}
\pdfoutput=1
\usepackage{graphicx,amsfonts,color,comment,amsmath,hyperref,float}
\usepackage[dvipsnames]{xcolor}
\usepackage{bigints}
\usepackage[normalem]{ulem}

\begin{document}
\title{Charm contribution to ultrahigh-energy neutrinos from newborn magnetars}

\date{\today}

\author{Jose Alonso Carpio}
\affiliation{Department of Physics, The Pennsylvania State University, University Park, Pennsylvania 16802, USA}
\affiliation{Department of Astronomy and Astrophysics, The Pennsylvania State University, University Park, Pennsylvania 16802, USA}
\affiliation{Center for Multimessenger Astrophysics, Institute for Gravitation and the Cosmos, The Pennsylvania State University, University Park, Pennsylvania 16802, USA}
\author{Kohta Murase}
\affiliation{Department of Physics, The Pennsylvania State University, University Park, Pennsylvania 16802, USA}
\affiliation{Department of Astronomy and Astrophysics, The Pennsylvania State University, University Park, Pennsylvania 16802, USA}
\affiliation{Center for Multimessenger Astrophysics, Institute for Gravitation and the Cosmos, The Pennsylvania State University, University Park, Pennsylvania 16802, USA}
\affiliation{Center for Gravitational Physics, Yukawa Institute for Theoretical Physics, Kyoto, Kyoto 606-8502, Japan}
\author{Mary Hall Reno}
\affiliation{Department of Physics and Astronomy, University of Iowa,
Iowa City, Iowa 52242, USA}
\author{Ina Sarcevic}
\affiliation{Department of Physics,
University of Arizona, Tucson, Arizona 85721, USA}
\affiliation{Department of Astronomy and Steward Observatory,
University of Arizona, Tucson, Arizona 85721, USA}
\author{Anna Stasto}
\affiliation{Department of Physics, The Pennsylvania State University, University Park, Pennsylvania 16802, USA}

\begin{abstract}
Newborn, strongly magnetized neutron stars (so-called magnetars) surrounded by their stellar or merger ejecta are expected to be sources of ultrahigh-energy neutrinos via decay of mesons produced in hadronic interactions of protons which are accelerated to ultrahigh energies by magnetic dissipation of the spindown energy. 
We show that not only pions and kaons but also charm hadrons, which are typically neglected due to their small production cross sections, can represent dominant contributions to neutrino fluence at ultrahigh energies, because of their short lifetimes, while the ultrahigh-energy neutrino fluence from pion and kaon production is suppressed at early times due to their significant cooling before their decay. 
We show that the next-generation detectors such as Probe Of Extreme Multi-Messenger Astrophysics (POEMMA), Giant Radio Array for Neutrino Detection (GRAND) and IceCube-Gen2 have a good chance of observing neutrinos, primarily originating from charm hadrons, from nearby magnetars. 
We also show that neutrinos from nearby magnetar-driven merger novae could be observed in the time interval between $10^2$ s and $10^3$ s, where the charm hadron contribution is dominant for neutrino energies above $10^8$ GeV, of relevance to next generation detectors. 
We also comment on potential impacts of the charm hadron contribution to the diffuse neutrino flux. 
\end{abstract}

\maketitle

\section{Introduction}
%
Magnetars are neutron stars with the highest magnetic fields in the Universe \cite{Harding:2006qn,Mereghetti:2008je,Kaspi:2017fwg}.
The magnetar birth rate is expected to be more than $\sim10$\% of the core-collapse supernova (SN) rate. 
While the fast-rotating magnetar event rate is unknown, it would not exceed the rate of Type Ibc SNe. For fast rotating magnetars with initial periods $P_i\sim 1$ ms at birth, one can expect magnetic fields as high as $10^{14}-10^{16}$ G, due to the amplification of the field by a dynamo mechanism \cite{Duncan:1992hi,Thompson:1993hn}, although a significant fraction of the magnetars may be explained by the fossil field hypothesis \cite{Ferrario:2006ib}. 

The rotational energy can be extracted in a form of the Poynting energy and fast-rotating neutron stars or magnetars may provide a site for efficient ultrahigh-energy cosmic-ray (UHECR) acceleration \cite{Arons:2002yj}. 
These cosmic rays do not escape the SN ejecta, rather lose their energy, through $pp$ and/or $p\gamma$ interactions, to secondary particles. 
For example, secondary particles such as charged pions are produced, which then decay to high-energy neutrinos. 
Previous studies on neutrino production in magnetars mainly considered pion decays, \cite{Murase:2009pg,Kotera:2011vs,Fang:2013vla,Fang:2015xhg} where the pions come from $pp$ and $p\gamma$ interactions (but see also Ref.~\cite{Fang:2017tla}). 
In this work, we focus on magnetar scenarios where $pp$ interactions dominate at early times and in addition to pion decays, we also consider kaon decays and charmed hadron decays as the sources of neutrinos. 
Typically, charm production in various astrophysical scenarios can be neglected because its production cross section is small when compared to pion/kaon production, however, it may not always be the case as considered for choked gamma-ray burst jets \cite{Enberg:2008jm,Bhattacharya:2014sta} (but see also Ref.~\cite{Murase:2013ffa}). 
In magnetars, for energies above $\mathcal{O}(10^{9})$ GeV, pions and kaons are subject to strong cooling due to interactions with surrounding protons and photons, while charmed hadrons decay promptly without any significant energy loss. If sufficient cooling is present, 
prompt decays of charm hadrons can dominate the neutrino fluence at ultrahigh energies. 
We show that charm hadron contributions become important in the context of next generation detectors, such as IceCube-Gen2 \cite{Aartsen:2019swn}, Probe Of Extreme Multi-Messenger Astrophysics (POEMMA) \cite{Olinto:2017xbi} and Giant Radio Array for Neurtino Detection (GRAND) 200k \cite{Alvarez-Muniz:2018bhp}, which are sensitive to $10^9$~GeV - $10^{11}$~GeV neutrinos. These detectors could potentially observe neutrinos from magnetars.

The organization of this paper is as follows. We first introduce the magnetar model and its UHECR injection spectrum, and then formulate hadron production and subsequent decay into neutrinos. 
We then present our results for a nearby magnetar, showing how charm contributions are dominant at the highest energies. We scan the magnetar parameter space to find where we can get significant charm contributions to the neutrino fluence. 
We also consider not only the magnetar-driven supernova scenario but also the magnetar-driven merger nova scenario, where charm contributes the most to the fluence at the highest energies, but where pion and kaon decays are still significant sources of neutrinos at lower energies. 
We evaluate the diffuse neutrino flux from these two scenarios. The appendices include details for charm production and the meson leptonic and semileptonic decay formulas used in our evaluations.

\section{Methods}\label{sec:methods}

The origin of UHECRs at energies of $\gtrsim10^{9.5}$~GeV has been an enigma for more than fifty years. 
The UHECRs around $10^{9}$~GeV are comprised mostly of protons, whereas at even higher energies the mass composition of UHECRs may be dominated by intermediate or heavy nuclei. However, there are significant uncertainties that come from hadronic interactions and UHECR measurements (see, e.g., Refs.~\cite{Anchordoqui:2018qom,AlvesBatista:2019tlv} for reviews). 

In what follows, we assume that the UHECRs accelerated in newborn magnetars are protons. See Ref.~\cite{Fang:2013vla} for discussion in the case of the nuclear composition. UHECRs from the newborn magnetars are mostly depleted unless the ejecta is punctured, so the maximum neutrino fluence is not much affected.
We evaluate neutrino injection rates from a given proton injection spectrum $dN_p/dE_p$ by calculating the particle spectra of the chain $pp\to hX \to \nu Y$, where $h$ is a hadron (pion, kaon and charmed hadron) that decays into neutrinos. The initial proton spectrum is a time-dependent function that depends on the magnetar's parameters,  e.g., the magnetic field, radius, initial period, moment of inertia, efficiency of acceleration and shock velocity, as discussed in the next section. The hadronic spectrum $dN_h/dE_h$ can be obtained from the initial proton spectrum at the source and the proton-proton differential cross section that gives $F_{pp\to h}$, the hadron production spectrum. 
The neutrino spectra are obtained by decaying these hadrons.

In this paper we make a distinction between the particle injection rates $dN/dE$ with dimensions of energy$^{-1}$time$^{-1}$ (the 
differential $dt$ is omitted) and the spectrum $F_{i\to f}$ of the final product of a single $i\to fX$ collision or decay. 
\subsection{Magnetar environment}
To determine the proton injection spectra, we start by considering properties of the magnetars and their mechanism for accelerating protons to high energies.  The magnetar consists of a rapidly rotating neutron star, with an initial angular frequency $\Omega_i = 2\pi/P_i$ and initial period $P_i\sim1$~ms. 
Neutron stars are known to spin down, and their rotational energy is carried by the wind, consisting of the outflowing plasma and magnetic fields. Charged particles are accelerated by tapping a fraction of the voltage available in the wind, via the wake-field acceleration mechanism \cite{Arons:2002yj}.

The spindown luminosity at time $t$ is given by
\begin{eqnarray}\nonumber
	L(t)&=&\frac{B_\text{NS}^2 R_\text{NS}^6\Omega_i^4}{4c^3}(1+\sin^2\chi)\left(1+\frac{t}{t_\text{sd}}\right)^{-2}\\\nonumber
	&\simeq& 1.5\times 10^{50}\text{erg s}^{-1}\, B_{\text{NS},15}^2R_{\text{NS},6}^6\Omega_{i,4}^4\\
	& & \times (1+t/t_\text{sd})^{-2} \;,
	\label{LuminosityEq}
\end{eqnarray}
where $\chi$ is the angle between the rotation and magnetic axes.  
Note that the above formula based on magnetohydrodynamics simulations \cite{Gruzinov:2004jc,Spitkovsky:2006np,Tchekhovskoy} is analogous to the well-known vacuum dipole formula. 
Our numerical values are obtained with $\langle \sin^2\chi\rangle=2/3$.
Throughout our work, for a quantity $Q$ we define $Q_x=Q/10^x$, where $Q$ is given in CGS units. The only exception to this convention are the ejecta masses $M$, where $M_{x} = M/M_\odot$.

From Eq.~\eqref{LuminosityEq}, it follows that for $t\gg t_\text{sd}$, the luminosity will decrease as $t^{-2}$ and does not depend on $\Omega_i$, since the spindown time $t_{\rm sd}$ depends on $\Omega_i^{-2}$.
In particular, for a neutron star with magnetic field $B_\text{NS}$, radius $R_\text{NS}$ and moment of inertia $I$, the spindown time is \cite{Murase16}
\begin{equation}
	t_\text{sd} =\frac{6I c^3}{5\Omega_i^2B_\text{NS}^2R_\text{NS}^6} \simeq 10^{2.5} \,\text{s}~\ I_{45}B_{\text{NS},15}^{-2}
	R_{\text{NS},6}^{-6}\Omega_\text{i,4}^{-2}.
\end{equation}

As noted above, we assume a proton composition of cosmic rays, and for simplicity, we assume that all accelerated protons at $t$ have a monotonic energy \cite{Arons:2002yj,Murase:2009pg} 
\begin{eqnarray}\nonumber
	E^M(t)&  = & \frac{f_\text{acc}eB_\text{NS}R_\text{NS}^3}{2c^2}\Omega_i^2\\\nonumber 
	& \simeq & 1.3\times 10^{13}~\text{GeV}\,  f_{\rm acc,-1}B_{\text{NS},15}R_{\text{NS},6}^{3}\Omega_{i,4}^2\\
	& & \times\left(1+t/t_\text{sd}\right)^{-1},
\label{EnergyEq}
\end{eqnarray}
where $f_{\rm acc}$ parametrizes the efficiency of the acceleration process. 

We assume that the proton injection rate is determined by the Goldreich-Julian rate~\cite{Goldreich:1969sb}, in which the proton injection rate spectrum at $t$ is written as
\begin{equation}
\frac{dN_p}{dE_p} = \frac{B_{\rm NS}R_{\rm NS}^3\Omega_i^2}{ec(1+t/t_{\rm sd})}\delta[E_p-E^M(t)],
\end{equation}
which roughly gives
\begin{eqnarray}\nonumber
\frac{dN_p}{dE_p} &\sim& 7\times 10^{39}~{\rm GeV}^{-1}~B_{\rm NS,15}R_{\rm NS,6}^3\Omega_{i,4}^2\\
& &(1+t/t_{\rm sd})^{-1}\delta[(E_p-E^M(t))/{\rm GeV}].
\end{eqnarray}
In the limit $t\gg t_\text{sd}$, we see that $E^M(t)\propto t^{-1}$ and is independent of $\Omega_i$. The time integration of $dN_p/dE_p$ gives a proton time-integrated injection spectrum that scales as $E_p^{-1}$~\cite{Murase:2009pg}.

\subsubsection{Magnetar-driven supernovae}
At the birth of the magnetar, the supernova ejecta propagates outward with speed $\beta_\text{ej}c$. We estimate the SN ejecta radius as 
\begin{equation}
r_\text{ej} \approx \beta_\text{ej}ct\simeq
10^{13.5}\ {\rm cm}\ \beta_{\text{ej},-1}t_4. 
\end{equation}
Note that, in general, $\beta_\text{ej}$ may depend on time, but we take it to be time independent for simplicity. 
The nucleon density in the ejecta is assumed to be homogeneous, such that $n_N = 3M_\text{ej}/(4\pi r_\text{ej}^3 m_p)$,
where $M_\text{ej}$ is the ejecta mass and $m_p$ is the proton mass. 
We may assume that the supernova ejecta masses may typically lie between $10M_\odot$ and $35M_\odot$ \cite{Heger:2004qp}.

\subsubsection{Magnetar-driven merger novae}
Rapidly rotating magnetars could be born at the merger of low-mass neutron star binaries.
At the merger, a significant amount of the mass would be ejected by dynamical interactions and/or disk winds. Typical ejecta masses lie in the range $10^{-2}M_\odot$-$10^{-1}M_\odot$ solar masses \cite{Wu:2016pnw}.

In the merger case, the rotational energy can be used to accelerate the ejecta. Thus, the ejecta speed $\beta_{\rm ej}$ is time dependent in general, and is found by solving
\begin{equation}
\Gamma_{\rm ej}(\beta_{\rm ej}) M_{\rm ej}c^2 = 
\Gamma_{\rm ej}(\beta_{\rm ej,0}) M_{\rm ej}c^2
+\int_0^t L(t) dt,
\end{equation}
where $\Gamma_{\rm ej}(\beta) = (1-\beta^2)^{-1/2}$ is the Lorentz factor of the ejecta. We then calculate the ejecta radius
\begin{equation}
r_{\rm ej}(t) = \int_0^t \beta(t')cdt', 
\label{eq:eq8}
\end{equation}
which determines the time-dependent nucleon density $n_N$.

\subsection{Hadronic spectrum}

The hadronic spectrum at the source depends on the
hadronic spectrum from a single $pp$ interaction, $F_{pp\to h}$, which is given by
\begin{equation}
F_{pp\to h}(E_h,E_p) = \frac{1}{E_p}\frac{1}{\sigma_{pp}(E_p)}\frac{d\sigma}{dx_E}(E_h,E_p),
\label{PPSpectrum}
\end{equation}
where $\sigma_{pp}(E_p)$ is the total inelastic $pp$ cross section, and $x_E=E_h/E_p$, where
$E_h$ is the hadron energy and $E_p$ is the proton energy in the lab frame. The time-dependent hadronic spectrum at the source is found via convolution of 
Eq. \eqref{PPSpectrum} with the proton injection rate:
\begin{equation}
	\frac{dN_h}{dE_h}(E_h) = \int_{E_h}^\infty dE_p \frac{dN_p}{dE_p} F_{pp\to h}(E_h,E_p).
\label{ppHadronSpectrum}
\end{equation}
We include $h=\pi,\ K$ and charm hadrons $h= D^0,D^{\pm},\ D_s,\ \Lambda_c$.

We use SIBYLL 2.3c \cite{Ahn09,Riehn17,Engel:2019dsg} to calculate the differential cross sections $d\sigma/dx_E$ for $pp\to \pi X$ and $pp\to KX$ 
interactions. Recent accelerator measurements, including ones from LHC at $\sqrt{s}=13$ TeV, have been used to improve hadronic interaction modeling in SIBYLL 2.3c compared to earlier versions.

For charmed hadron production, we use the relation between differential energy distributions of the charm quark and charmed hadron
(see, e.g., Ref. \cite{Bhattacharya15,Bhattacharya16}),
\begin{equation}
\label{eq:dsdx}
\frac{d\sigma}{dx_E}(x_E,E_p)=\int_{x_E}^1\frac{dz}{z} \frac{d\sigma}{dx_c}(x_c,E_p) D_c^h(z),
\end{equation}
where $x_c=E_c/E_p=x_E/z$, $d\sigma/dx_c$ is the $pN\to cX$ production cross section and $D_c^h$ is the fragmentation function. The quantity $x_E$ translates to the hadron energy by $E_h=x_E E_p$.
We use the fragmentation function $D_c^h$ of Kniehl and Kramer
\cite{Kniehl:2006mw}.  This fragmentation function was also used, for example, in the evaluation of the prompt atmospheric neutrino flux from charm in ref. \cite{Bhattacharya15,Bhattacharya16}. The fragmentation function includes the corresponding fragmentation fractions for charm quarks to fragment into $D^+$, $D^0$, $D_s^+$ and $\Lambda_c^+$, equal to the fragmentation functions for antiquarks to the corresponding antiparticle hadrons \cite{Lisovyi:2015uqa}.

There are large uncertainties in the theoretical predictions of hadronic production of charm. The strong interaction corrections depend on powers of the strong coupling constant, evaluated at characteristic energy scales comparable to the mass of the produced quark. The charm quark mass $m_c$, taken here to be $1.3$ GeV, means that the theoretical uncertainties are large, even in the next-to-leading order (NLO) QCD evaluation in the collinear parton model 
\cite{Nason:1987xz,Nason:1989zy,Mangano:1991jk}. 

Another source of uncertainty at high energies is the fact that the neutrino fluence from charm contributions depends on the small momentum fraction (small-$x$) parton distribution functions (PDFs), especially the gluon PDF, 
 of relevance to evaluating $d\sigma/dx_E$ in Eq. (\ref{PPSpectrum}). 
At these high energies, the values of $x$ probed in the $pp$ interactions are extremely small, beyond the range which was probed in the high energy accelerators. Thus the PDFs are largely unconstrained in this region and need to be extrapolated.
One theoretical approach to small-$x$ PDFs is the $k_T$-factorization framework \cite{Catani:1990eg,Collins:1991ty,Levin:1991ya,Ryskin:1995sj}
in its linear formulation that accounts for resummation of 
large logarithms $\ln(1/x)$, and in its non-linear formulation that also accounts for saturation effects \cite{Gribov:1984tu} of the gluon density at very small-$x$.  

In the results shown here, we perform a NLO QCD evaluation of $d\sigma/dx_c$ in the collinear approach. This is our central result, which is in reasonable agreement with SIBYLL after fragmentation is included. We also evaluate the differential cross section for charm production in the  $k_T$ factorization framework with linear and non-linear evolution of the gluon PDF density. Details are included in appendix \ref{app:charm}. The result is that the span of predictions is a factor of $\sim 1/3 - 3$ of the central NLO QCD curve for most of the range of $x_E$ values. This factor of $1/3-3$ uncertainty is represented by the shaded blue band in the results from charm shown below. We discuss the evaluation of the charmed hadron contribution in more detail in appendix \ref{app:charm}.

Inside the ejecta, hadrons will interact with the ambient protons, leading to hadronic cooling. Since the magnetic field in the SN shock
is weak, we neglect synchrotron losses. 
We account for hadronic cooling by comparing the cooling timescale $t_\text{cl}$
to the decay timescale $t_\text{dec}^h = E_h\tau_h/m_h$, where $\tau_h$ is the lifetime of the hadron. For example, the pion
cooling timescale is given by $t_\text{cl}\approx t_{\pi N}\approx (\kappa_{\pi p}\sigma_{\pi p} n_N c)^{-1}$, where $\sigma_{\pi p}$ is the pion-proton inelastic cross section and $\kappa_{\pi p}$ is the average inelasticity, and $n_N$ is the nucleon density (defined in Section II. A. 1). We can then modify the hadronic injection rate with a cooling factor $1-\exp(-t_{\rm cl}/t^h_{\rm dec}$). Analogous expressions are obtained for kaon and charmed hadron 
cooling timescales. The $\pi p$, $Kp$ inelastic cross sections are obtained from SIBYLL, while the charmed hadron-proton cross sections are
assumed to be equal to the $Kp$ cross section.
The typical energy range of interest is $10^{10}$~GeV-$10^{12}$~GeV.
The inelasticities are assumed to be energy-independent, with $\kappa_{h p}=0.8$ for all hadron-proton interactions, including for charmed hadrons.

The hadronic injection rates are also modified by 
the effective optical depth of the $pp$ reaction, which is $f_{pp}\approx\kappa_{pp}\sigma_{pp}n_Nr_\text{ej}\simeq
5.7\times 10^4\,  M_{\text{ej},1}\beta_{\text{ej},-1}^{-2}t_4^{-2}$ for $\sigma_{pp}\sim 10^{-25}\, \text{cm}^{-2}$ and $\kappa_{pp}\sim 0.5$ in the case of a magnetar driven supernova.
The modification of the hadronic injection rate is thus
\begin{equation}
\label{eq:cooling}
\frac{dN_h}{dE_h}(E_h) \longrightarrow f_{pp}\frac{dN_h}{dE_h}(E_h)\left[1-\exp\left(-\frac{t_\text{cl}}{t_\text{dec}^h}\right)\right]\ .
\end{equation}
Cooling in the merger case is treated in a similar fashion.  Secondary pion production from $\pi p$ interactions are neglected, which can affect the spectra by a factor $\mathcal{O}(1)$ at earlier times \cite{Murase:2009pg}.


\subsection{Hadronic decays}

We turn next to hadronic decays into neutrinos.
The neutrino spectra from two-body pion and kaon leptonic decays are well documented (see, e.g., Ref. \cite{Lipari93}). The  time-dependent neutrino spectrum from hadronic lepton or semi-leptonic decays is of the form 
\begin{equation}
\frac{dN}{dE_\nu}(E_\nu) = \int dE_h \frac{dN_h}{dE_h} F_{h\to \nu}(E_\nu,E_h),
\label{eq:twobody}
\end{equation}
where $F_{h\to \nu}(E_\nu,E_h)$ is the decay spectrum of the neutrino from hadron $h$. The details for
$h=\pi,\ K$ are collected in appendix \ref{app:piondecays}, and those for charm decays are in appendix
\ref{app:charmdecays}.

For pions, only the two-body decays are relevant. Leptonic decays of kaons are included, but 
contributions from other kaon
decay modes, namely the hadronic mode $K^+\to\pi^+\pi^0$ followed by $\pi^+\to \nu_\mu \mu^+$, and the semileptonic decay mode $K^+\to\pi^0e^+\nu_e$ are neglected.
The hadronic kaon decay mode contribution to the neutrino spectrum is suppressed by cooling factors.
The suppression of secondary decay contributions in the context of the atmospheric lepton flux is illustrated in, for example, Ref.~\cite{Pasquali:1997cg}.
Contributions from the semileptonic charged kaon decay modes are suppressed by the branching ratio. 

Muons accompany muon neutrinos in the decays. We include $\mu\to \nu_\mu$ contributions given
the muon spectrum. In charged pion decays, most of the energy is carried by the muon. 
The differential neutrino spectrum from the $h\to\mu\to\nu$ chain, in the absence of cooling, is
\begin{eqnarray}
\frac{dN_\nu}{dE_\nu}(E_\nu) &=& \int_{E_\nu}^\infty dE_\mu \int_{E_\mu}^{E_h^{\rm max}} dE_h \frac{dN_h}{dE_h}(E_h)\\
\nonumber
& \times & F_{h\to\mu}(E_\mu, E_h)\,  F_{\mu\to\nu}(E_\nu,E_\mu)\ .
\end{eqnarray}
For the case when $h=\pi$, 
$F_{\pi\to\mu}$ is the muon distribution from pion decay which depends on the energy of pion, and
$F_{\mu\to\nu}$ is the neutrino distribution from muon decay which depends on muon energy. The maximum hadron energy $E_h^{\rm max}$ is provided in Appendix B.

In two-body decays like $\pi\to \mu \nu$, the 
neutrino spectrum from $\mu\to \nu$ depends on the muon polarization $h_{\pi\to \mu}$ in $F_{\mu\to\nu}$, which in turn depends on the pion energy. To account for hadronic cooling in $\pi\to\mu\to\nu$, we find the average polarization $\langle h_{\pi\to \mu}\rangle$ using the pion spectrum at production.
Muons will also cool in the supernova due to $\mu$ collisions with nucleons, characterized by a cooling timescale 
$t_\text{cl}\approx (\sigma_{\mu N}^{\rm eff}n_N c)^{-1}$, where $\sigma_{\mu N}^{\rm eff}$ is the effective cross section that includes energy losses due to pair production, bremsstrahlung, and muon-nuclear interactions. It also depends on the ejecta composition, and we take  
$\sigma_{\mu N}^{\rm eff}\sim10^{-29}$ cm$^2$, likely on the larger size than the realistic value. But our results are insensitive to our choice of $\sigma_{\mu N}^{\rm eff}$, affecting fluence contributions at early times only, where cooling is significant.
To account for this cooling,  we include a cooling factor 
of $(1-\exp(-t_{\rm cl}/t_{\rm dec}^\mu))$ that multiplies the muon spectrum at production.  Detailed formulas are included in appendix \ref{app:piondecays}.

\begin{figure}[t]
\includegraphics[width=0.5\textwidth]{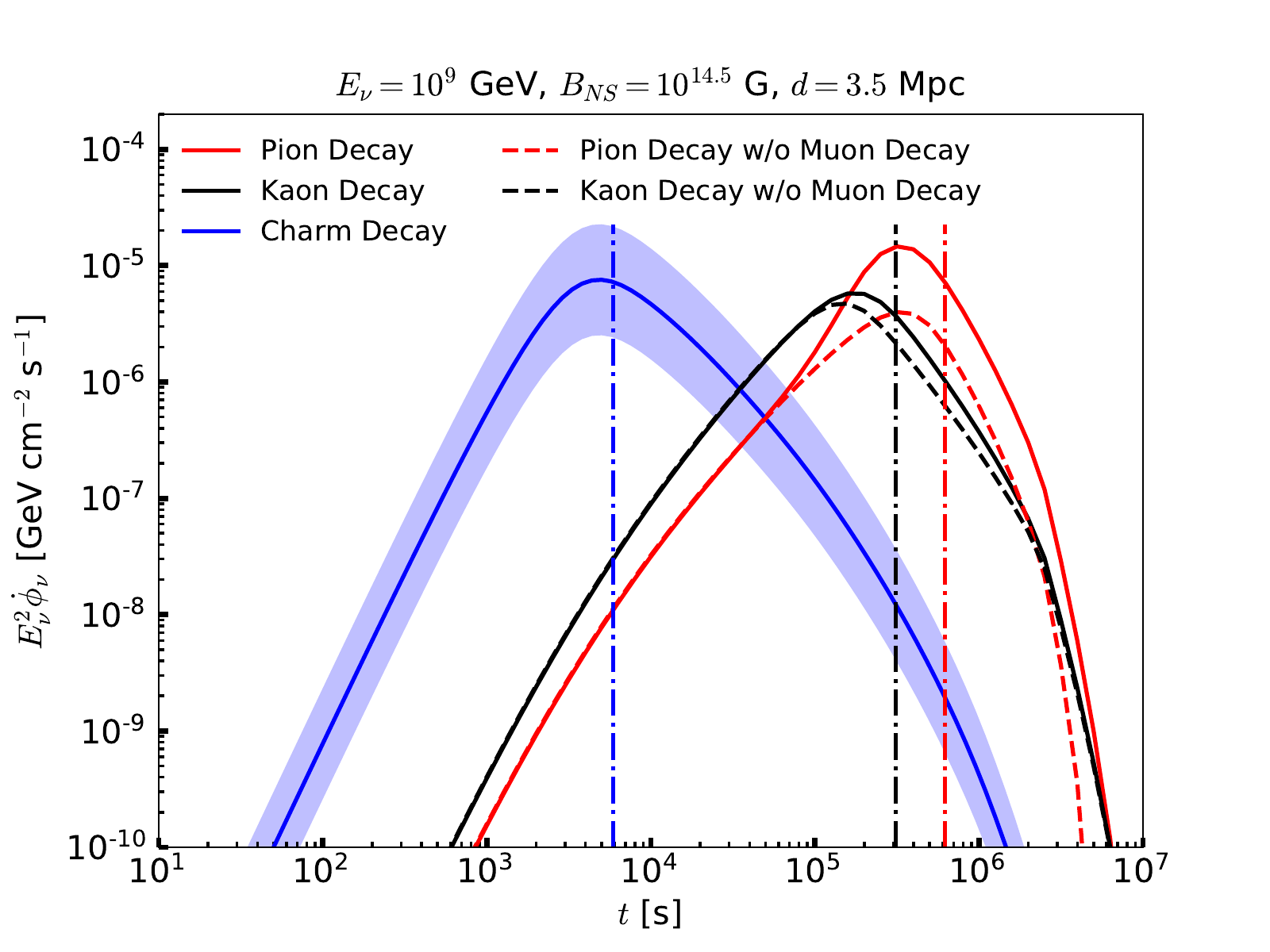}
    \vspace{-20pt}
\caption{All-flavor neutrino light curve $E_\nu^2\dot{\phi}_\nu$ at $E_\nu=10^9$ GeV of a magnetar at a distance of 3.5 Mpc. The charm uncertainty factor of 1/3--3 around the central curve is given by the shaded blue region. For the case of neutrinos from pion (kaon) decay, we include an additional dashed red (black) curve to isolate the $\nu_\mu$ component from $\pi^+\, (K^+) \to \mu^+ \nu_\mu$ and charge conjugate, without taking into account the contributions from the muon decay. The dot-dashed vertical lines indicate the locations where decay time and cooling time are equal, based on our estimate given by Eq. \eqref{CritEstimate}. Here, the spin-down time is $t_{\rm sd}=10^{3.5}$ s.}
\label{LightCurve}
\vspace{-10pt}
\end{figure}

For charmed hadron decays, we use an effective neutrino energy distribution approximating them by three-body decays
\cite{Bhattacharya16,Bugaev98}. The semi-leptonic charmed hadron decay channels also include pions, kaons and muons that decay to neutrinos, but their
contributions to the overall neutrino spectrum can be neglected compared to the contributions from $pp\to \pi/K\to \nu$ processes 
because the 
charmed hadron production cross section is already small compared to the pion production cross section, and in addition, 
any mesons or 
muons coming
from charmed hadron decays experience significant cooling. 

\section{Results}
\subsection{Magnetar-driven supernovae}\label{sec:sn}

Our first magnetar model assumes $M_\text{ej}=10~M_\odot$, $\beta_\text{ej}=0.1$, $I=10^{45}\ \text{g cm}^{2}$, $B_\text{NS} = 10^{14.5}\ \text{G}$, $R_\text{NS}=10^6\ \text{cm}$ and $f_{\rm acc}=0.1$. The initial angular frequency is $\Omega_i=10^4$ s$^{-1}$, an optimistic value because its corresponding period $P_i=0.6$ ms is close to the minimum period of a neutron star \cite{Haensel:1999mi}. The associated spindown time is $t_\text{sd}=10^{3.5} \ \text{s}$.  For the purposes of observation estimates, we consider a nearby magnetar at a distance of $d=3.5$~Mpc. We calculate the neutrino injection rate $dN_\nu/dE_\nu$ and convert it to the observed single source flux $\dot{\phi}_\nu = (1/4\pi d^2)
dN_\nu/dE_\nu$, where $d$ is the source distance. 

The neutrino optical depth can be estimated as
$\tau_{\nu N}= n_N\sigma_{\nu p} r_\text{ej}\simeq3.1\times 10^{-2} M_\text{ej,1}\beta_\text{ej,-1}^{-2}t_4^{-2}$, where we take $\sigma_{\nu p}\sim 10^{-32}$ cm$^{-2}$, which is the order of magnitude for the neutrino-nucleon charged current cross section in the $10^9$ GeV - $10^{10}$ GeV range.
We can thus neglect neutrino attenuation effects in our calculations, except for $t<10^3$ s. Fortunately, for such early times, the flux does not significantly contribute to the 
fluence. 
For this work, we consider the all-flavor neutrino flux, so neutrino oscillation effects are ignored. While we won't separate the fluxes from a single source by flavor, we will separate it into  its source components from pion, kaon and charm, namely, $\dot{\phi}_{\nu,\pi},\dot{\phi}_{\nu,K}$ 
and $\dot{\phi}_{\nu,c}$ respectively. Similar notations will be used when referring to the fluence and the diffuse neutrino flux. 

We show the neutrino light curves from pions, kaons and charm at $E_\nu=10^9$ GeV in Fig. \ref{LightCurve}. Here we observe the expected pattern of neutrinos from charm decay dominating first, followed by contributions of kaon and pion decay at later times.
We also found that $D^0$ decays contribute the most to the 
charm component of the neutrino flux, due to its larger production
cross section compared to other charmed hadrons.

We see that for $t<t_\text{sd}$, all fluxes are suppressed. In this regime, where we can assume $L(t)$ and $E^M(t)$ are time independent, the time dependence is carried by the cooling factor. The large matter density leads to a short cooling time, and the cooling factor is well approximated by $t_{\rm cl}/t_{\rm dec}\propto t^3$, where the $t^3$ power law comes from the $n_N\propto t^{-3}$. The charm flux peaks slightly after $t_\text{sd}$, when cooling time and decay time are equal, and will continuously decrease afterward, as a result of the luminosity decrease. The time dependence of the spectral function $F_{pp\to h}$ due to its dependence on $E_p$ will also mildly contribute to the flux suppression (see Eq. \ref{ppHadronSpectrum}).

In the case of the kaon and pion components of the neutrino flux, the impact of the luminosity decrease is not as significant as the exponential increase in the cooling factor, causing the $t^3$ behavior to shift to an approximate $t^2$ power law above $t_\text{sd}$. We stress that this tail of the pion component of the light curve, and its $t^2$ dependence, can be significantly modified by secondary pion production and cause a flatter light curve. 
At $t\sim 3\times 10^5$ s, pion and kaon cooling stops, and the flux suppression is caused by the luminosity decrease. At $t\sim 5\times 10^6$ s there is a sharp cutoff that is caused by the corresponding cutoff in the pion/kaon flux due to $E_{\pi/K}/E_p$ approaching unity, as well as the decrease in the efficiency $f_{pp}$. A similar effect occurs for the charm component, but is not shown in the figure.  

Above $t\sim 10^5$ s, we observe a small bump in the pion flux, which is caused by the muon decay component that is no longer suppressed by its corresponding muon cooling factor (see dashed red line in Fig. \ref{LightCurve}). This feature is less prominent with kaons; contributions from muon decays in this case are much smaller because they come from the  high $y_K=E_\mu/E_K$ region, where the distribution function is much smaller.

In Fig. \ref{LightCurve}, for our parameter choices
and $E_\nu=10^9$ GeV, coincidentally, pion, kaon and charm sources of the fluxes have approximately the same $E^2\dot{\phi}$ peak. The peak flux is approximately determined by the critical time,  when $t_\text{dec}=t_\text{cl}$. 
Because of the different $t$ dependence of the fluxes, at higher (lower) neutrino energies, the pion and kaon peak fluxes are lower (higher) than the charm peak flux.

However, we note that the relative positions of the peaks, for any neutrino energy, is the same, since it depends on the ratios of the 
hadron masses and their lifetimes, neglecting small difference in the energy dependence of the cross section, $\sigma_{h p}$ for different hardon. The exception to this rule occurs when the neutrino energy is close to the proton energy.

The critical time, $t_h^{cr}$, which is the time at which the decay time, $\tau_h E_h / m_hc^2$,  is equal to the cooling time, 
$t_\text{cl}\approx ( \kappa_{hp}\sigma_{h p}^{}n_N c)^{-1}$, is given by
\begin{equation}
t_h^{\rm cr} \simeq 68\; \text{s} \left(\frac{E_h}{m_hc^2}\right)^{1/3}
M_{\rm{ej},1}^{1/3}\sigma_{hp,-25}^{1/3}\beta_{\rm{ej},-1}^{-1}
\tau_{h,-9}^{1/3}.
\label{CritEstimate}
\end{equation}

A slight deviation from this relationship is present in our simulations because of the inherent time dependence of the hadron-proton cross section
$\sigma_{hp}$. We can estimate the critical energy $E_h^{cr}$ at which cooling time is equal to decay time. We use Eq. \eqref{CritEstimate}, substituting $t_h^{\rm cr}$ with $t$ and $E_h$ with $E_h^{\rm cr}$ and solving for $E_h^{\rm cr}$.
The estimated value of $E_h^{\rm cr}$ increases with time.

We applied these estimates to Fig. \ref{LightCurve} and marked
these critical times with dot-dashed lines. To convert the 
neutrino energy $E_\nu=10^9$ GeV to a hadron energy, we estimate
$E_h=4E_\nu$ for $h=\pi,K$ and $E_h=3E_\nu$ for $D^0$. Eq.
\eqref{CritEstimate} somewhat overestimates the time at which the peak flux occurs because
of the simplified form of the equation, which does not include time dependence of the luminosity, for example.

\begin{figure}
    \centering
    \includegraphics[width=0.5\textwidth]{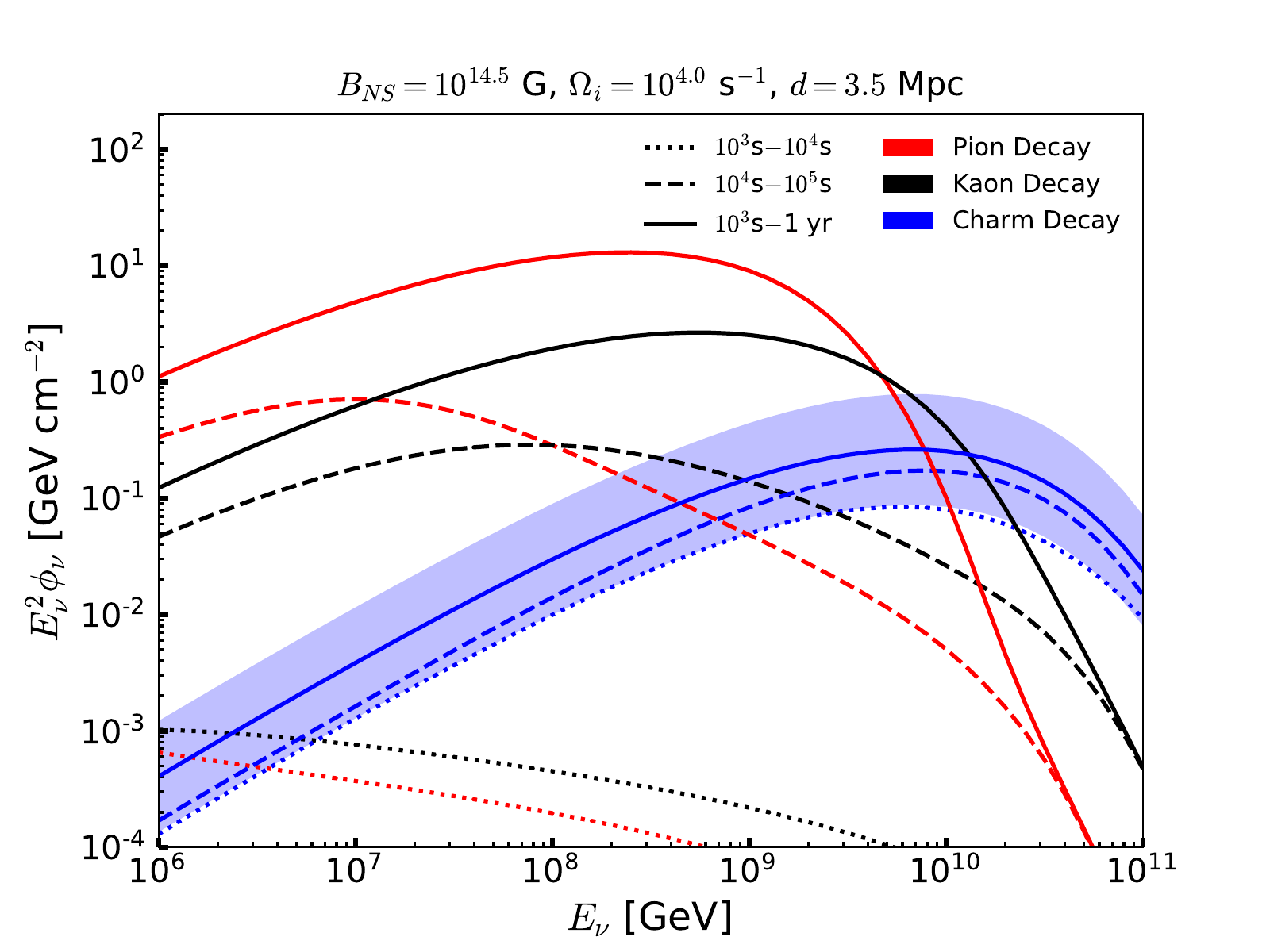}
    \vspace{-20pt}
    \caption{All-flavor fluence of high-energy neutrinos of a nearby magnetar at a distance of 3.5 Mpc, for different time intervals.
A band in the charm spectrum in the time interval $10^3$~s -- $1$~yr is shown, spanning a factor of $1/3-3$ times the central result.}
    \label{FluenceNoSensitivity}
    \vspace{-10pt}
\end{figure}

\begin{figure}
\centering
\includegraphics[width=0.5\textwidth]{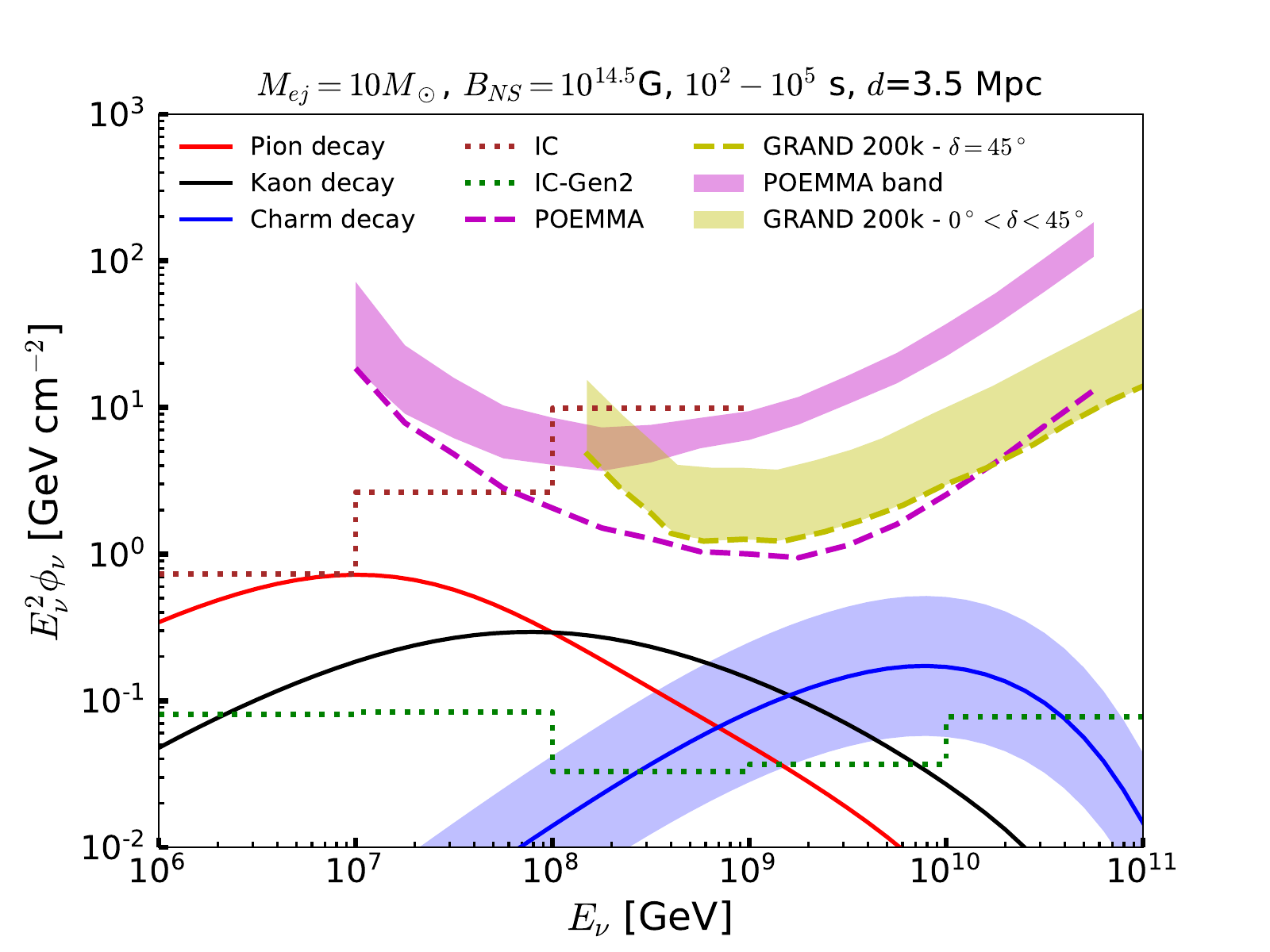}
\vspace{-20pt}
\caption{Neutrino fluence in the interval $10^2-10^5$~s compared to the long burst sensitivities of various experiments. A band in the charm spectrum is shown, spanning a factor of $1/3-3$ times the central result.
The IceCube 90\% CL upper limit on the spectral fluence from GW170817 on a 14-day window \cite{Aartsen:2019swn} (dotted brown line), while the IceCube-Gen2 curve is the 90\% sensitivity for an event at a similar position in the sky \cite{Aartsen:2019swn} (dotted green line). The best 90\% unified CL sensitivity per energy decade for long bursts for POEMMA is given by the dashed purple line, while its the purple band is the sensitivity range over most portions of the sky \cite{Venters:2019xwi}. The 90\% CL sensitivity for GRAND 200K in the optimistic case of a source at declination $\delta=45^\circ$ is shown by the dashed yellow line, and the yellow band is the declination-averaged sensitivity $0^\circ <\delta < 45^\circ$ \cite{Alvarez-Muniz:2018bhp}. 
}
\label{FluenceSupernova}
\vspace{-10pt}
\end{figure}

The all-flavor fluence, $\phi_\nu$,
scaled by neutrino energy squared for the model is shown in Fig.
\ref{FluenceNoSensitivity}, for three time intervals. 
We observe that the pattern in Fig. \ref{LightCurve} extends over a wide energy range: pion and kaon fluxes are suppressed below $10^4$ s, when the neutrino flux is predominantly from charm decay. The most energetic protons are accelerated at early times, where strong hadronic cooling of pions and kaons occurs. Consequently, the neutrino flux is dominated by charm decay at the highest energies, followed by kaon decay and finally pion decay, in order of their respective decay times. Unlike the neutrino light curves, the time dependent proton energy cutoff effects are not seen in Fig. \ref{FluenceNoSensitivity} as they get smeared out by the time integration, with the exception of the absolute cutoff given by $E^M(t=0)$, which lies outside the chosen energy range. 

For  $t>10^5$ s, the proton number density is very low and cooling effects become negligible. At late times, we see pion contributions dominating the neutrino fluence, with a maximum
of $E^2\phi_\nu$ at $10^9$ GeV  for this model. 
We expect the maximum value of $E^2\phi_\nu$ from the pion contribution to be at a lower energy than that from charm because proton injection energy decreases with time. 

\begin{figure*}[t]
\centerline{
\includegraphics[width=0.5\textwidth]{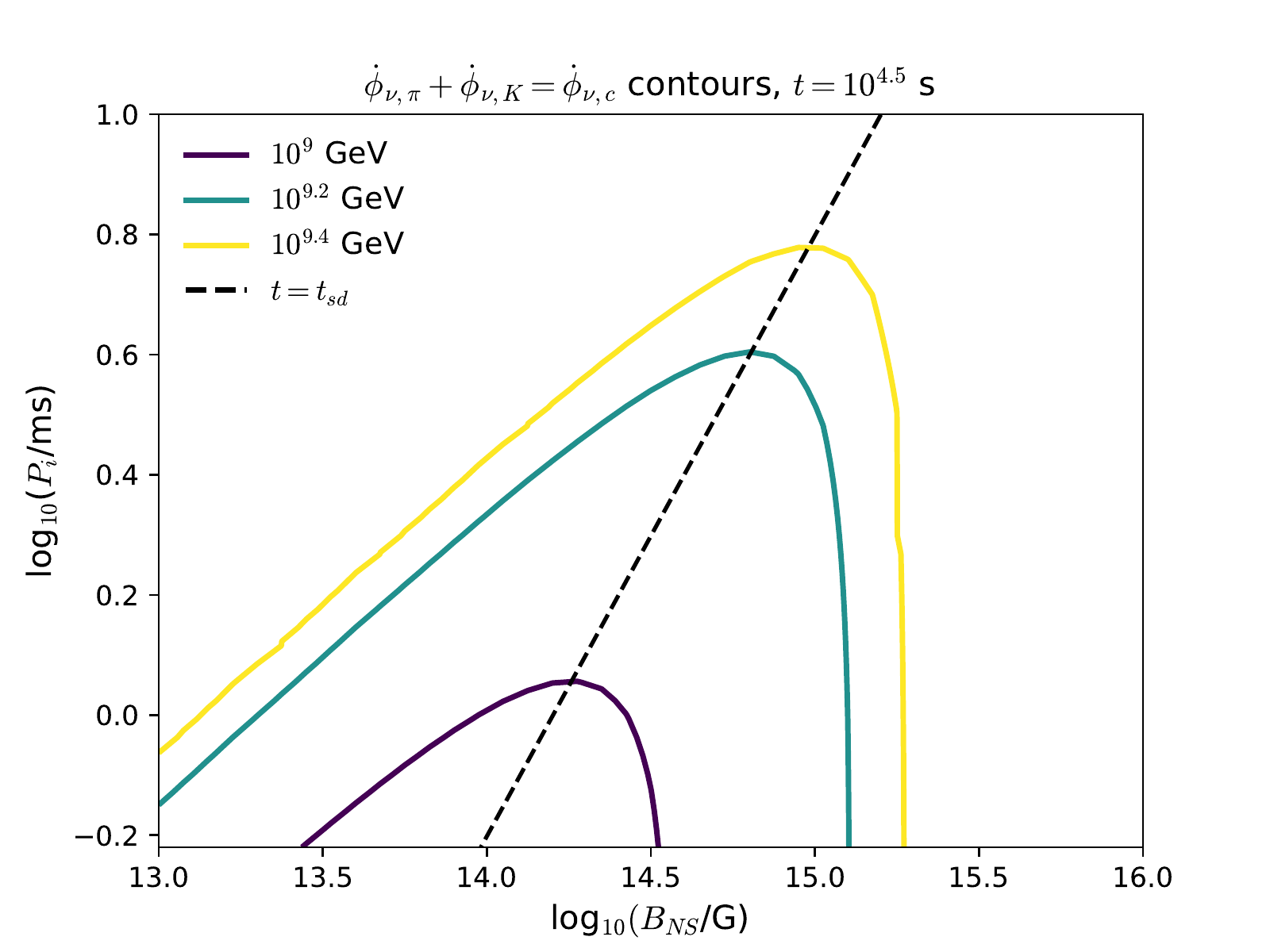}
\includegraphics[width=0.5\textwidth]{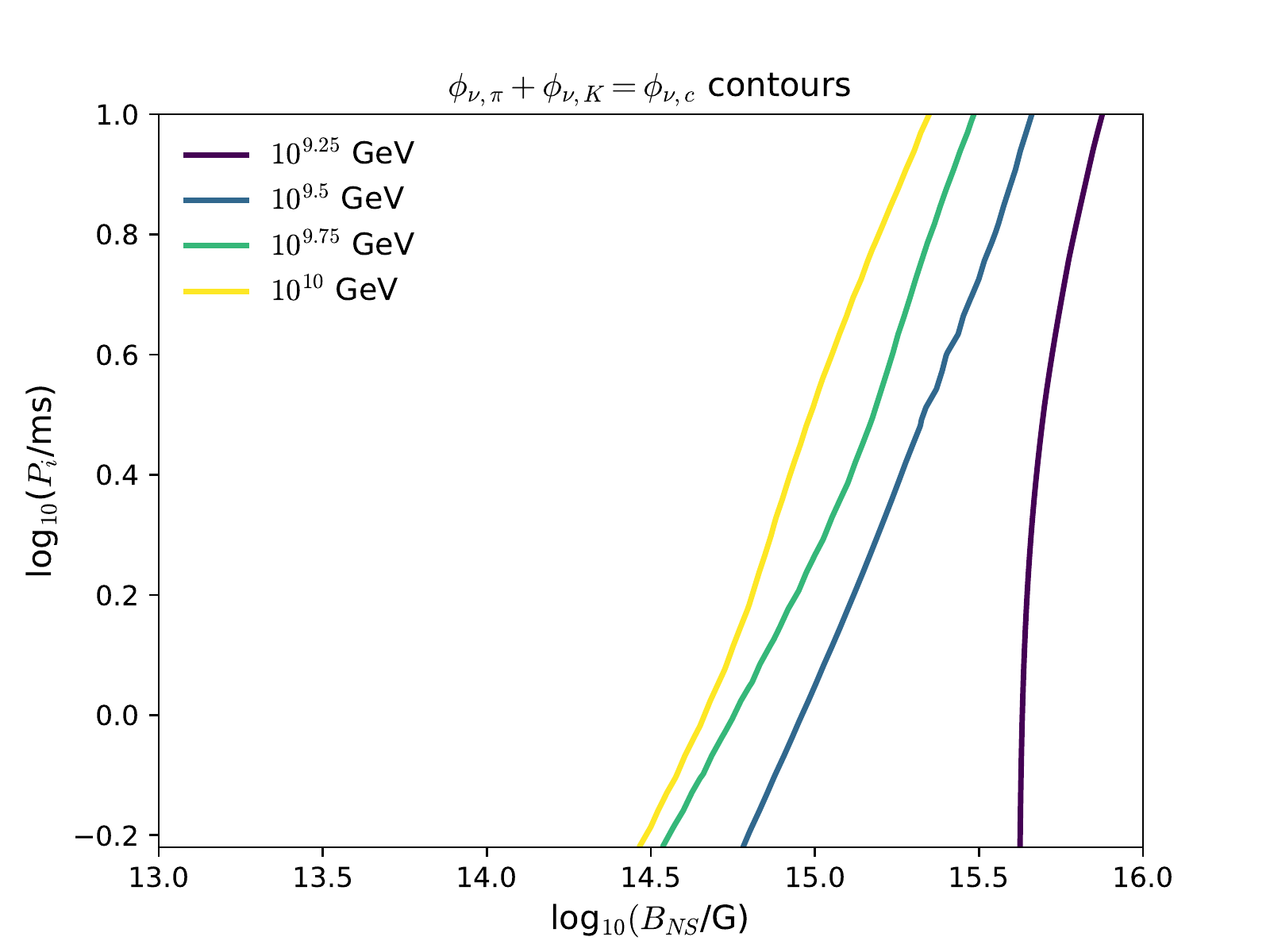}
}
\vspace{-10pt}
\caption{Left panel: Contour plots where $\dot{\phi}_{\nu,\pi}+\dot\phi_{\nu,K} = \dot\phi_{\nu,c}$ at the injection time $t=10^{4.5}$ s. We have also marked the line where $t=t_\text{sd}$. The lower limit in the period corresponds to the minimum spin period of a neutron star, $P_i\sim 0.6$ ms \cite{Haensel:1999mi}. The parameter space below the solid curves
have $\dot{\phi}_{\nu,c}> \dot{\phi}_{\nu,\pi}+\dot{\phi}_{\nu,K}$ for a given energy.
Right panel: Same as left panel, but using the total fluence $\phi$ instead of the flux $\dot\phi$ at a fixed time. The parameter space to the right of the solid curves have
$\phi_{\nu,c}> \phi_{\nu,\pi}+\phi_{\nu,K}$ for a given energy.}
\label{ParameterSpaceScan}
\vspace{-15pt}
\end{figure*}

We also varied the mass $M_{\rm ej}$ to 20$M_\odot$ and $30M_\odot$ and compared the fluences with those of Fig. \ref{FluenceNoSensitivity}. In the time interval $10^3-10^4$ s, all fluxes are suppressed by approximately the same factor above $10^9$ GeV, while charm is not very sensitive to $M_{\rm ej}$ below this energy. At larger times, fluence become less sensitive to mass. This insensitivity manifests itself at lower energies first, where decay time is shorter. The total fluence is very insensitive to mass, and the fluence does not vary more than a factor $\sim 2-3$ because late time emissions contribute the most to the total fluence, when the cooling time is very large.

A separate all-flavor fluence calculation was made with $B_\text{NS}=10^{15}$ G (other parameters remain the same). Our results for $t>t_\text{sd}$ are in agreement with those of Ref.~\cite{Murase:2009pg}, which use the same parameter set. 

In Fig. \ref{FluenceSupernova} we compare the neutrino fluence, in the interval $10^2$ s - $10^5$ s, with the sensitivities of various experiments to a long burst. We show the IceCube 90\% CL upper limit on the spectral fluence from GW170817 in a 14 day window \cite{Aartsen:2019swn} to illustrate IceCube's current sensitivity. The projected 90\% CL sensitivities for IceCube-Gen2 for a similar position in the sky (green dotted histogram) \cite{Aartsen:2019swn}, for POEMMA's best case scenario (purple dashed curve) and sensitivity range over most portions of the sky (purple
band) \cite{Venters:2019xwi}, and for GRAND 200K for the declination average over $0^\circ<\delta<45^\circ$ (yellow band) and for
$\delta = 45^\circ$ (dashed yellow curve) \cite{Alvarez-Muniz:2018bhp} are also shown.

We see in Fig. \ref{FluenceSupernova} that the pion component can be detected in IceCube in the 1 PeV - 10 PeV range, but the kaon component will be below the sensitivity curve for this model. IceCube Gen-2, however, would pick up all the components above $10^8$ GeV. If we have a magnetar at a distance of 1 Mpc, POEMMA and GRAND 200K can detect the charm component, although such an event would be rare.  In the case of a shorter burst of less than $10^3$ s, where the POEMMA and GRAND 200K sensitivities are better, we find that the fluence is not large enough to reach these sensitivities. 

We also studied the parameter sets where we can get significant charm contributions. To do this, we look at the $B_{\rm NS} - P_i$ parameter space, keeping all other parameters listed at the beginning of the section fixed. For each 
($B_{\rm NS},P_i$) pair, we look at
the energy where $\phi_{\nu, \pi}+\phi_{\nu, K}=\phi_{\nu, c}$, that is, the energy where the neutrino flux from pions and kaons falls below the neutrinos from charm. We first look at these contours for $t=10^{4.5}$ s, which are shown in the left panel of Fig. \ref{ParameterSpaceScan}. With a fixed injection time, the cooling factors depend primarily on the Lorentz factor $E_h/m_h$ because the hadron-proton inelastic cross section grows slowly with energy. Thus, the proton energy becomes the relevant variable when scanning the parameter space, as this determines the hadronic spectrum. The region of $B_{\rm NS} - P_i$ parameter space below the solid curves have $\dot{\phi}_{\nu,c}> \dot{\phi}_{\nu,\pi}+\dot{\phi}_{\nu,K}$ for a given energy.
The diagonal black dashed line in the left panel of Fig. \ref{ParameterSpaceScan} shows $t=t_\text{sd}$. To the left of the black dashed line, the luminosity $L(t)$ is constant and to the right, it is proportional to $t^{-2}$ (see Eq. \eqref{LuminosityEq}). As mentioned above, for $t\gg t_\text{sd}$ the proton energy $E^M(t)$ becomes independent of $P_i$, which is why we get the vertical lines on the contours. 

In the right panel of Fig. \ref{ParameterSpaceScan}, we make a similar study using the total fluence, {where
the region of $B_{\rm NS} - P_i$ parameter space to the right of the solid curves have $\Phi_{\nu,c}> \Phi_{\nu,\pi}+\Phi_{\nu,K}$ for a given energy.}
When comparing fluence, the value of the $t_\text{sd}$ is important: strong magnetic fields and small values of $P_i$ are preferred, as this increases the proton energy and enhances the charm spectrum. Late time emission is dominated by pion and kaon contributions, when $t\gg t_{\rm sd}$. It follows that these fluences depend on $B_{\rm NS}$, but are independent of $P_i$ (see Eq. \eqref{LuminosityEq} and Eq. \eqref{EnergyEq}). On the other hand, charm contributions depend on both parameters, where smaller $P_i$ increases the proton energy and the luminosity at early times, where charm production is relevant.

We emphasize that if the spindown time falls below $10^2$ s, the neutrinos need to come from early decays, however, at early times, the proton density is high enough to cool even the charm hadrons. In addition, if the spindown time is small as a result of a large $B_\text{NS}$, the luminosity will be much lower at later times because $L(t)\propto B_\text{NS}^{-2}t^{-2}$ for $t\gg t_\text{sd}$ (see Eq. \eqref{LuminosityEq}). We thus find that, while stronger $B_\text{NS}$ is preferred to get a charm dominated flux at the highest energies, such a choice would hinder our ability to detect the neutrino flux.

\begin{figure}[t]
    \centering
    \includegraphics[width=0.5\textwidth]{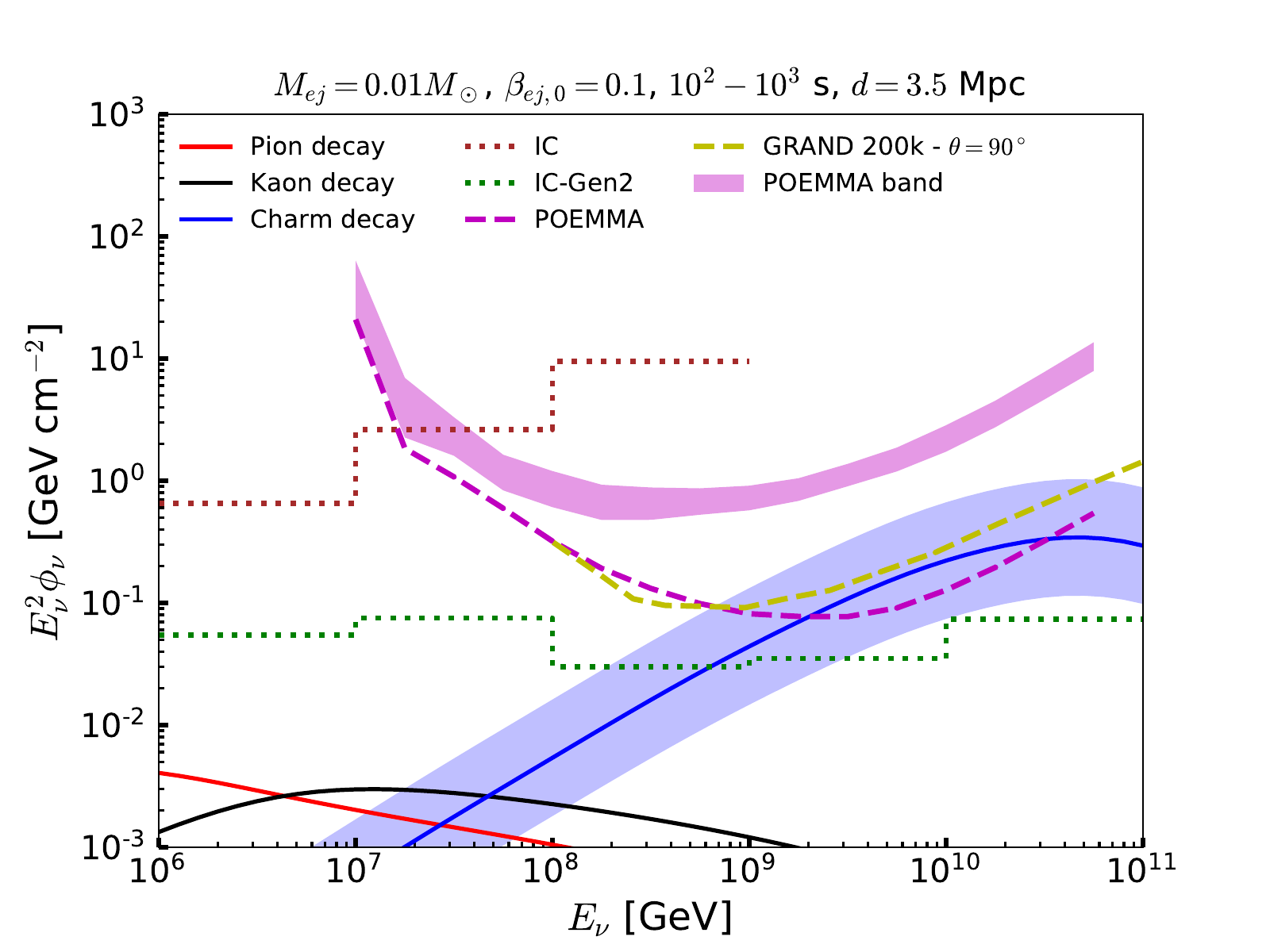}
    \vspace{-20pt}
    \caption{Neutrino fluence of a nearby neutron star merger at a distance of 3.5 Mpc, in the interval $10^2-10^3$ s, compared to the short burst sensitivities of various experiments. The IceCube 90\% CL upper limit on the spectral fluence from GW170817 on a $\pm 500$ s time window \cite{Aartsen:2019swn} is shown with a dotted brown line, while the IceCube-Gen2 curve is the 90\% sensitivity for an event at a similar position in the sky \cite{Aartsen:2019swn} (dotted green line). The best 90\% unified CL sensitivity per energy decade for short bursts for POEMMA is given by the dashed purple line, while its the purple band is the sensitivity range over most portions of the sky \cite{Venters:2019xwi}. The 90\% CL sensitivity for GRAND 200K in the optimistic case of a source at zenith angle $\theta=90^\circ$ is shown by the dashed yellow line \cite{Alvarez-Muniz:2018bhp}.}
    \label{SensitivityShort}
    \vspace{-20pt}
\end{figure}

\subsection{Magnetar-driven merger novae}
Another scenario of interest is neutrino production from merger ejecta. We use $I=10^{45}~\text{g cm}^{2}, B_\text{NS} = 10^{15}~\text{G}, R_\text{NS}=10^6~\text{cm}$, $f_\text{acc}=0.1$ and $\Omega_i=10^4$~s$^{-1}$.
For the ejecta mass, we use $M_{\rm ej}=0.01~M_\odot$ and initial speed $\beta_{\rm ej,0}=0.1$. Changing the ejecta mass by a factor of 2 has negligible impact on the fluence.

The ejecta is less massive than the magnetar case and its speed increases with time, so cooling effects are weaker. This allows for enhanced neutrino production at earlier times, because charm hadrons will decay before cooling. We see in Fig. \ref{SensitivityShort} that, for a nearby merger, next generation experiments could see the charm component, within a 1000 s time window, for sources optimally located for detection. The pion and kaon components, on the other hand, are suppressed below the sensitivity curve and would only be observable at later times.

\section{Discussion}

\subsection{Diffuse neutrino intensity}


\begin{figure*}[t]
\centerline{
\includegraphics[width=0.5\textwidth]{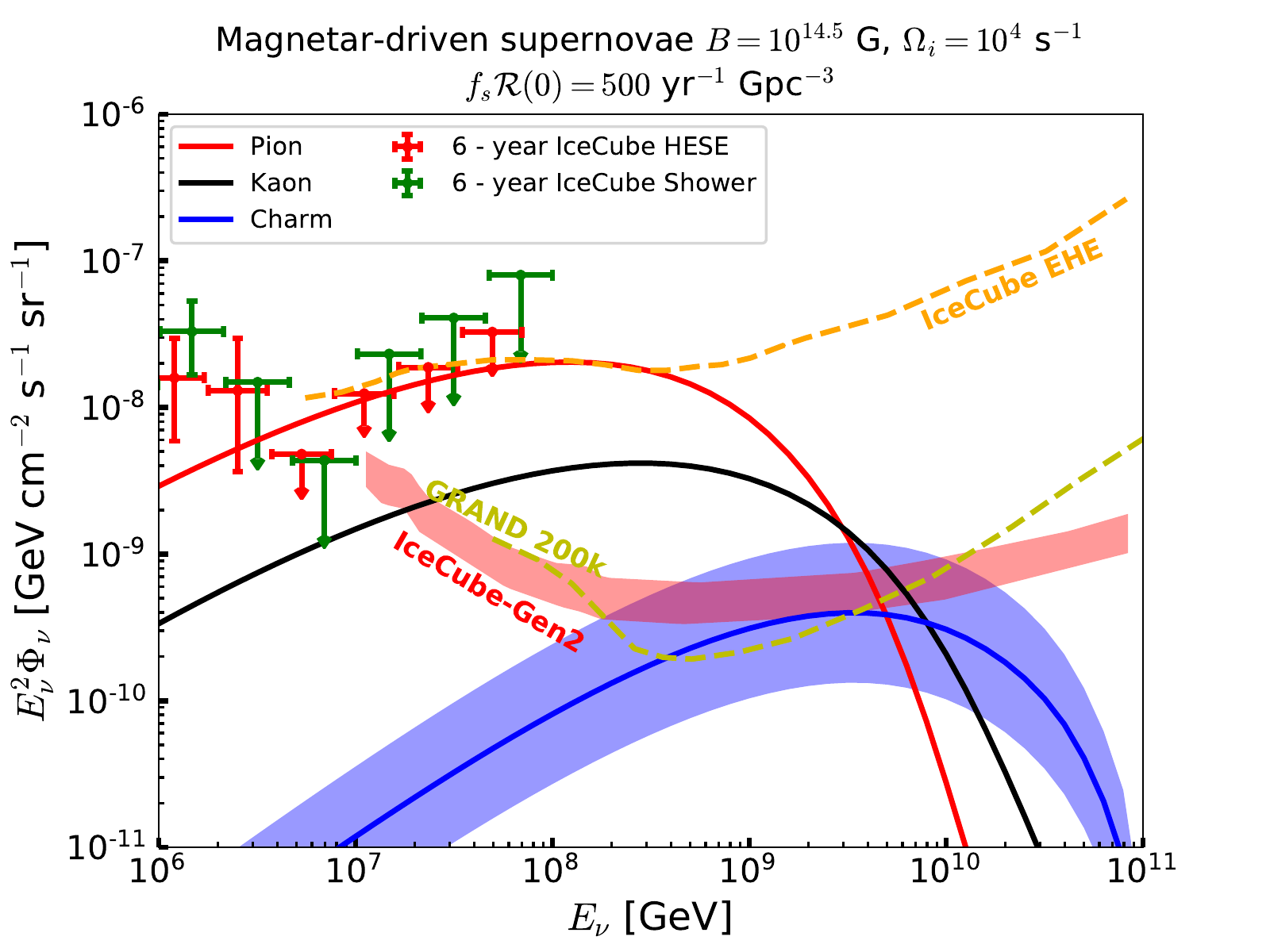}
\includegraphics[width=0.5\textwidth]{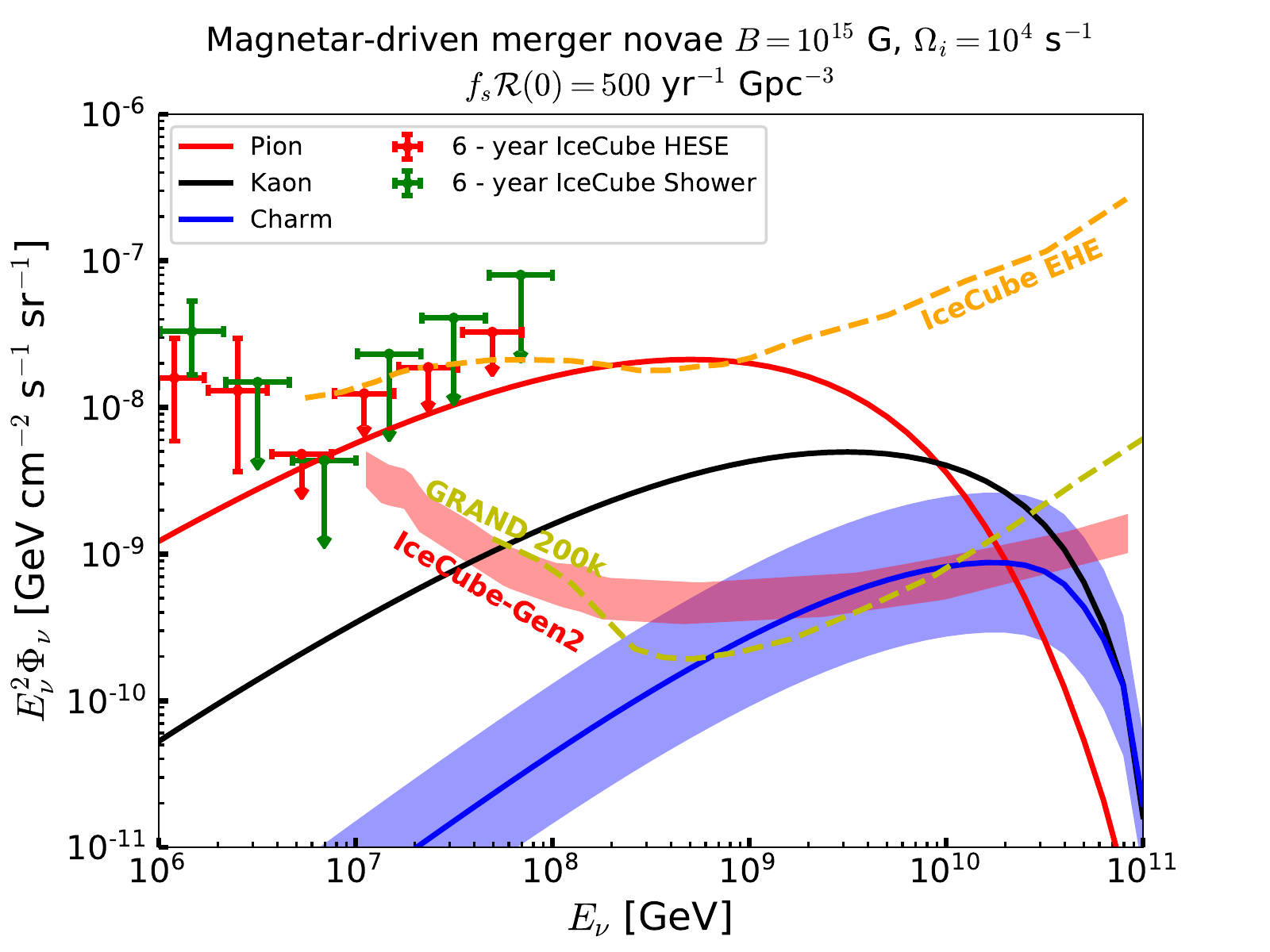}
}
\vspace{-10pt}
\caption{Left panel: Magnetar-driven supernovae contributions to the all-flavor diffuse neutrino flux. The red error bars show the results of the IceCube 6-year HESE analysis, obtained by multiplying the per-flavor neutrino flux in Ref.~\cite{Kopper:2017zzm} by a factor of 3. The green error bars correspond to the IceCube 6-year shower analysis \cite{Aartsen:2020aqd}. The 5-year IceCube-Gen2 sensitivity is shown by the red band \cite{Aartsen:2019swn}, while the 10-year GRAND200k sensitivity is shown by the yellow curve and is scaled from the 3-year sensitivity \cite{Alvarez-Muniz:2018bhp}. The orange curve is the IceCube nine-year 90\% CL EHE diffuse flux upper limit \cite{Aartsen:2018vtx}. Right panel: Same as the left panel, showing instead magnetar-driven merger novae contributions to the diffuse neutrino flux.}
\label{DiffuseFlux}
\vspace{-10pt}
\end{figure*} 

The sources discussed in Section III will also contribute to the diffuse neutrino flux. The corresponding all-flavor diffuse neutrino flux, $\Phi_{\nu}$ is given by
\begin{equation}
\Phi_\nu=\frac{cf_s}{4\pi}\int_0^{z_{\rm max}}\mathcal{R}(z)\frac{dN}{dE'}[E(1+z)](1+z)\left|\frac{dt}{dz} \right| dz
\end{equation}
where $\mathcal{R}(z)$ is the local rate density of magnetar sources, $f_s \mathcal{R}(0)$, is a free 
parameter and its functional form can be parametrized as \cite{Madau:2014bja}
\begin{equation}
\mathcal{R}(z)=\mathcal{R}(0)(1+z)^{2.7}
\frac{1+[1/2.9]^{5.6}}{1+[(1+z)/2.9]^{5.6}}\ .
\end{equation}
We take $f_s\mathcal{R}(0)=500$ yr$^{-1}$ Gpc$^{-3}$ in our evaluation below. This value is consistent with observations as long as not all of the supernova and merger events are bright.
The prefactor $f_s$ takes into account effects from pair loading, particle acceleration mechanisms and other phenomena that could affect flux normalization.
The derivative $|dt/dz|$ is
\begin{equation}
    \left|\frac{dt}{dz}\right| = \frac{1}{H_0(1+z)\sqrt{\Omega_M(1+z)^3+\Omega_\Lambda}},
\end{equation}
with $\Omega_M=0.3$, $\Omega_\Lambda=0.7$ and $H_0=67$ km s$^{-1}$ Mpc$^{-1}$.

Contributions to the diffuse flux are shown in Fig.~\ref{DiffuseFlux}. We include the results from the IceCube six-year HESE \cite{Kopper:2017zzm} and six-year shower \cite{Aartsen:2020aqd} analyses. The diffuse flux sensitivities for IceCube-Gen2 (5-year) \cite{Aartsen:2019swn} and GRAND 200K (10-year) \cite{Alvarez-Muniz:2018bhp} are shown by the red band and yellow curve, respectively, while the IceCube extremely-high-energy (EHE) diffuse flux upper limit (9-year) is shown by the orange curve. 
For both magnetar-driven supernovae and merger novae, we see that charm decay does not significantly contribute to the diffuse flux, because the flux is dominated by pion decay at late times. 
The fluxes for both types of supernovae and merger novae can remain below current IceCube limits if the rate is $f_s\mathcal{R}(0)=500$ yr$^{-1}$~Gpc$^{-3}$, and next-generation detectors can see the pion component up to $E_\nu \sim10^{10}$~GeV. In the case of merger novae, cooling at early times is not as strong as the supernova case. At times $t>10^4$~s, pions and kaons will decay before cooling, and will contribute significantly to the fluence, even at the highest energies. 
For magnetar-driven supernovae, the separation between charm and pion components is more pronounced, but the diffuse flux from charm hadron decay is not sufficiently high to be detected by IceCube-Gen2.

We point out that, for both scenarios, there is some tension between our models and the IceCube EHE limits, because the model dependent limits would be more stringent than the differential limit shown in Fig.~\ref{DiffuseFlux} \cite{Aartsen:2018vtx}. However, given model uncertainties such as the local rate density, one can evade these constraints.

\subsection{Effects of the photomeson production}
One of the possible caveats of this work is that we ignore the photomeson production. 
Details are model dependent and in principle depend on two kinds of radiation fields. 
One is radiation thermalized in the ejecta, while the other is thermal or nonthermal radiation from the wind bubble. 
If the radiation is thermal, the ejecta temperature is estimated to be $kT\approx0.4~{\rm keV}~{\mathcal E}_{\rm rad,51}^{1/4}{(\beta_{\rm ej}/0.1)}^{-3/4}{(t/1000~{\rm s})}^{-3/4}$, where ${\mathcal E}_{\rm rad}$ is the radiation energy. The threshold photomeson production is $E_p \sim 0.2\ {\rm GeV}^2/(3kT)\sim 0.2\times{10}^{6}~{\rm GeV}~{\mathcal E}_{\rm rad,51}^{-1/4}{(\beta_{\rm ej}/0.1)}^{3/4}{(t/1000~{\rm s})}^{3/4}$, which is typically lower than the proton energy given by Eq.~(\ref{EnergyEq}).
Above the threshold, the photomeson production optical depth is approximately given by Ref.~\cite{Murase:2009pg}
\begin{eqnarray}
f_{p\gamma}&\approx&\kappa_{p\gamma}\sigma_{p\gamma}n_\gamma R_{\rm ej}\nonumber\\
&\simeq&380~{({\mathcal E}_{\rm rad}/{10}^{51}~{\rm erg})}^{3/4}{(\beta_{\rm ej}/0.1)}^{-5/4}\nonumber \\
&\times &{(t/10^4~{\rm s})}^{-5/4},
\end{eqnarray}
where $\kappa_{p\gamma}\sim0.2$ is the inelasticity and $\sigma_{p\gamma}$ is the photomeson production cross section. Note that the multipion production is important in the case of the thermal radiation field. 
This can be compared to the effective $pp$ optical depth, which is given by
\begin{eqnarray}
f_{pp}&\approx&\kappa_{pp}\sigma_{pp}n_{\rm ej} R_{\rm ej}\nonumber\\
&\simeq&5.7\times{10}^4~{({M}_{\rm ej}/10~M_{\odot})}{(\beta_{\rm ej}/0.1)}^{-2} \nonumber\\
&\times & {(t/{10}^4~{\rm s})}^{-2},
\end{eqnarray}
where $\kappa_{pp}\sim0.5$ is the inelasticity and $\sigma_{pp}$ is the $pp$ cross section.
Thus, as long as energy injected by the central engine is thermalized, interactions with baryonic matter are more important at early times. 
The transition occurs at
\begin{equation}
t_{\rm tr}\sim8\times{10}^6~{\rm s}~{({M}_{\rm ej}/10~M_{\odot})}^{4/3}{({\mathcal E}_{\rm rad}/{10}^{51}~{\rm erg})}{(\beta_{\rm ej}/0.1)}^{-1}.
\end{equation}
This implies that our results on the charm contribution are unlikely to be affected even if the thermal radiation field is included. This is because energy losses due to inelastic $pp$ collisions are dominant in the early phase during which the charm contribution is dominant at the highest energies.     

In addition, nonthermal particles may be generated at the termination shock inside the magnetar nebula (e.g., Refs.~\cite{Kotera:2013yaa,Murase:2014bfa,Kashiyama:2015eua}). 
Analogous to the Crab pulsar wind nebula, a significant fraction of the Poynting energy could be dissipated. If this is the case, the thermalization in the nebula matters, which could happen if the nebular Thomson optical depth satisfies $\tau_T^{\rm nb}\gtrsim\beta_{\rm nb}^{-1}$, i.e., $t\gtrsim2\times10^{4}~{\rm s}~M_{\rm nb,-7}^{1/2}\beta_{\rm nb,-1}^{-1/2}$, where $M_{\rm nb}$ is the nebular mass and $\beta_{\rm nb}$ is the nebular velocity. 
For example, in the merger case, this can happen if almost all the spindown energy is dissipated with the production of electron-positron pairs (see Ref.~\cite{Fang:2017tla} for such a case). Then, the model would need to be adjusted to include contributions from $p\gamma$ interactions, where charmed hadrons are not produced. 
However, such a situation can be realized only if the nebula is compact, in which the most of the thermalization occurs in the ejecta. 
Details depend on the magnetization and pair-loading of the wind that are uncertain. Also, if only a fraction of the spindown energy is dissipated in the nebula~\cite{Murase:2013mpa}, our assumptions can be justified. Note that our setup for the calculations is similar to those in the previous works~\cite{Fang:2013vla,Fang:2015xhg}. See Fig.~1 of Refs.~\cite{Murase:2009pg} for effects of the photomeson production (see also Ref.~\cite{,Fang:2017tla} for the merger case). 

\section{Conclusions}
We presented a study of ultrahigh-energy neutrino production by newborn magnetars, accounting for pion, kaon and charmed hadron production from $pp$ interactions in the supernova and merger ejecta. 
The charm component was obtained in the QCD calculation at NLO accuracy, together with an uncertainty band, a factor of $1/3-3$ around the NLO QCD flux that encloses the results obtained from 
$k_T$ factorization approaches and the SIBYLL Monte Carlo simulations.
The evolution of the proton injection spectrum and the ejecta expansion was included in the calculations, as well as the energy dependence of the various production cross sections. 
Using a benchmark parameter set, we found that for neutrino energies above $10^9$ GeV, charm contributions are much higher than the pion and kaon contributions at early times because hadronic cooling suppresses the neutrino fluxes from these latter contributions. When $t>t_\text{sd}$, the relative importance of kaon contributions increases as the ejecta's proton density decreases, followed by the pion contributions, in line with our expectations based on their lifetimes. The highest energies, above $10^{10}$ GeV, are dominated by charm contributions, essentially independent of pion/kaon contributions, and come from the most energetic protons which are injected at times $t<t_\text{sd}$.

We found that for $B_\text{NS}=10^{14.5}$ G and $P_i = 2\pi \times 10^{-4}$ s,
IceCube-Gen2 is projected to be sensitive to the charm component of the all-flavor neutrino fluence from a nearby magnetar at a distance
$\sim$ 3.5 Mpc, for locations such as that of GW17081. 
POEMMA and GRAND200k would be sensitive to such an event if it was located at a distance of $\sim$ 1 Mpc.  For the benchmark magnetar parameters, the accompanying pion and kaon contributions to the neutrino fluence at energies below $10^9$ GeV could also be observed by IceCube (pion only) and IceCube Gen-2 (both).

We scanned the $P_i-B_\text{NS}$ parameter space, to see at what energy the charm contributions to the neutrino flux overcome those of pions and kaons. Stronger magnetic fields and shorter periods are preferred, as this increases the proton energy at early times. However, these choices reduce the spindown time and cause cooling of charm hadrons, reducing their flux contributions below POEMMA and GRAND sensitivity curves.

In the case of a nearby neutron star merger, we found that neutrinos from charm hadron decay are likely to be observed by next generation detectors, within a time window of $\sim 1000$ s, without the accompanying lower energy neutrinos from pion and kaon decays. Both magnetar-driven supernovae and merger novae neutrino fluxes are consistent with IceCube's diffuse flux measurements.

Newborn magnetars have been expected to be the promising sources of gravitational waves, which is especially the case in the merger scenario (see Ref.~\cite{Bartos:2012vd} for a review). Even for the magnetar-driven supernova case, gravitational waves from a nearby event may be detected by current and future detectors if a magentar is deformed and/or subject to instabilities~\cite{Kashiyama:2015eua}. 
Our model demonstrates that newborn magnetars are interesting targets for multimessenger searches with gravitational waves and ultrahigh-energy neutrinos, as well as electromagnectic waves.

\vspace{-10pt}
\acknowledgements
We thank T. DeYoung and M. Ackermann for discussions of IceCube-Gen2 projected sensitivities. We also thank F. Riehn for helping us with SIBYLL.
J.C. is supported by the Fermi GI program
111180. The work of K.M. is supported by the Alfred P. Sloan Foundation, NSF Grant No.~AST-1908689, and KAKENHI No.~20H01901. M.H.R. is supported in part by the U.S. Department of Energy Grant DE-SC-0010113, I.S. is supported in part by the U.S. Department of Energy Grant DE-SC-000913 and A.M.S. is supported in part by the U.S. Department of Energy Grant DE-SC-0002145. A.M.S. is supported in part by National Science Centre in Poland, Grant 2019/33/B/ST2/02588.

\appendix


\section{Charm production} 
\label{app:charm}
In this Appendix, we include some of the details of our evaluation of charm production. The spectrum of neutrinos from charm hadron decays begins with the  energy distribution of these hadrons in $pp$ collisions, $d\sigma/dx_E$ to evaluate $F_{pp\to h}$ in Eq.~(\ref{PPSpectrum}). The charm quark distribution is evaluated in NLO QCD collinear approach \cite{Nason:1987xz,Nason:1989zy,Mangano:1991jk}, and with the $k_T$ factorization formalism \cite{Catani:1990eg,Collins:1991ty,Levin:1991ya,Ryskin:1995sj}. 
 In the latter case, calculation was based on the approach developed in \cite{Martin:2003us,Bhattacharya16} and two scenarios were considered for the evolution of the unintegrated parton density, the linear case as well as the non-linear case which includes corrections due to the large parton density \cite{Kutak:2003bd}. 
The unintegrated densities from \cite{Kutak:2012rf} were used, which were fitted to the inclusive HERA data. The charm quark distribution is then fragmented using fragmentation functions $D_c^h$. In Eq.~(\ref{eq:dsdx}), the fragmentation function $D_c^h$ used is that of Kniehl and Kramer~\cite{Kniehl:2006mw}, 
\begin{equation}
D_c^h(z) = \frac{Nz(1-z)^2}{[(1-z)^2+\epsilon z]^2}\, ,
\end{equation}
with the fit parameters given in Table \ref{Table1}. This  parametrization of the fragmentation functions is also used in the evaluation of the prompt atmospheric neutrino flux from charm in ref. \cite{Bhattacharya15,Bhattacharya16}. The constant $N$ for each hadron $H$ includes the fragmentation fractions $f_H$ for each particle \cite{Lisovyi:2015uqa}:
$f_{D^0}=0.606$, $f_{D^+}=0.244$, $f_{D_s^+}=0.081$ and $f_{\Lambda_c^+}=0.061$.

\setlength{\tabcolsep}{1.2em}
\begin{table}[h]\begin{center}
\begin{tabular}{ccc}
\hline\hline
	Particle & $N$ & $\epsilon$\\\hline
	$D^0$ & 0.577 & 0.101\\
	$D^+$ & 0.238 & 0.104\\
	$D_s^+$ & 0.0327 & 0.0322\\
	$\Lambda_c^+$ & 0.0067 & 0.00418\\
\hline\hline
\end{tabular}
	\caption{Parameters for the charm quark fragmentation function \cite{Kniehl:2006mw}. The factor $N$ is scaled to reproduce the fragmentation fractions of Ref. \cite{Lisovyi:2015uqa}.}
	\label{Table1}
\end{center}\vspace{-10pt}\end{table}

\begin{figure}[h]
\includegraphics[width=0.45\textwidth]{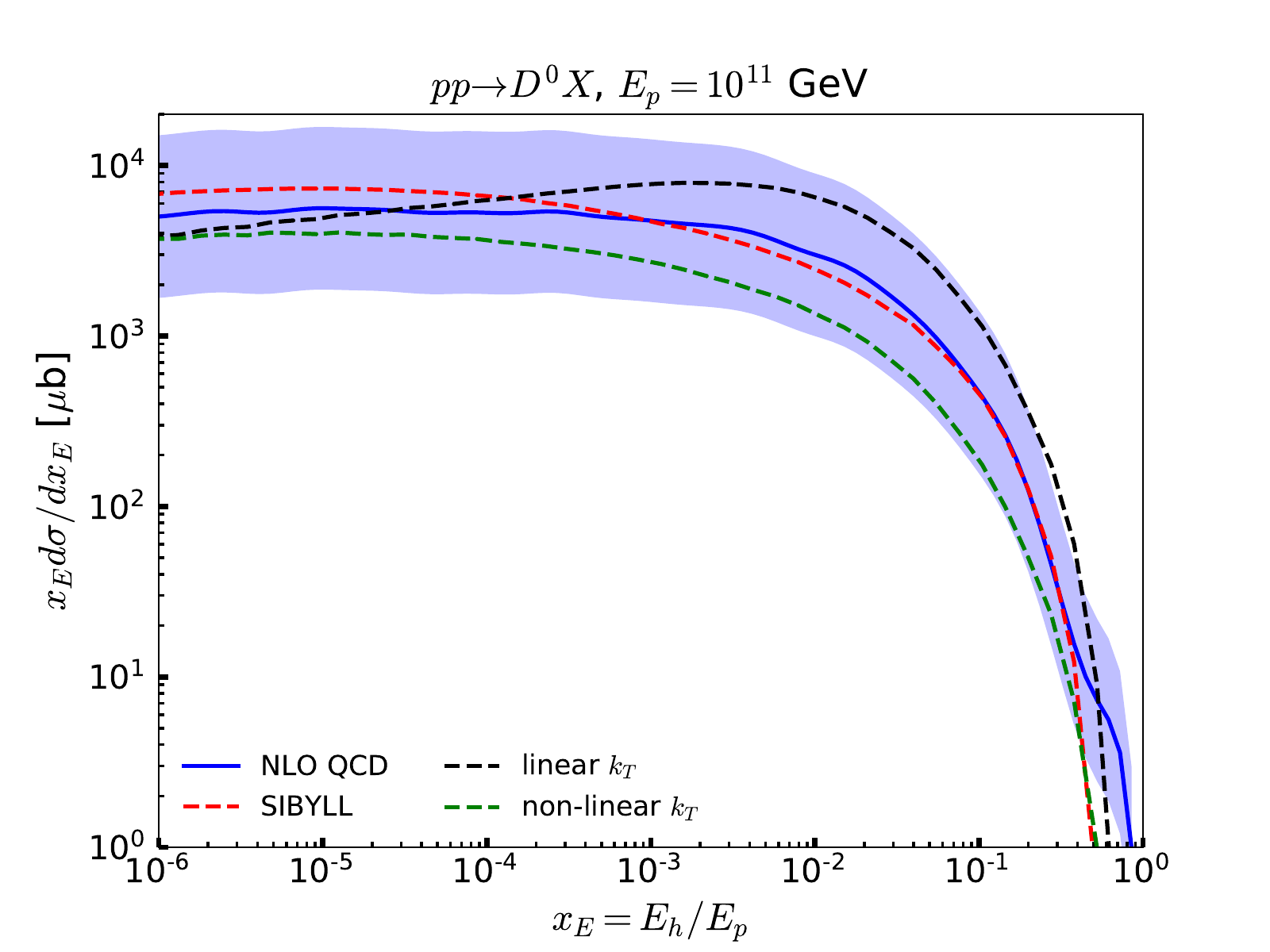}
\caption{As a function of $x_E=E_{D^0}/E_p$, the differential distribution of $D^0$ mesons produced in collisions of protons with $E_p$
incident of fixed target protons for $E_p=10^{11}$ GeV. The four curves show the evaluation using NLO QCD, the linear  and non-linear $k_T$ formulations and the SIBYLL result. The blue band spans a factor of $1/3-3$ times the NLO QCD result.}
\label{fig:charm-ep}
\end{figure}

As a representative case, in Fig. \ref{fig:charm-ep} the 
distribution $x_E d\sigma/dx_E$ for $x_E=E_{D^0}/E_p$ for three QCD approaches: linear and non-linear $k_T$
factorization and NLO QCD collinear  calculation 
for protons with energy $E_p=10^{11}$ GeV incident on a fixed proton target. Also shown are the SIBYLL 2.3c $x_E$ distributions for the $D^0$. The blue band shows a factor of $1/3-3$ of the NLO QCD collinear   result, representative of the range of theoretical uncertainties in the prediction for charm meson production. The band includes the predictions from the other three approaches, except at very large $x_E$.  There, the predictions differ more, and the small parton-$x$ extrapolation of the parton distribution functions in the collinear parton model show an effect. This very large $x_E$ region does not make a significant contribution to the neutrino fluence. The blue uncertainty band in Fig. \ref{fig:charm-ep} is translated to neutrino fluence calculations from charm production and decay.

\section{Meson leptonic decay formulas}
\label{app:piondecays}

For completeness,
we include in this appendix the basic equations for the pion decay chain \cite{Lipari93,Barr:1989ru,Gaisser16}.
The leptonic kaon decay chain formulas are identical,
after
substituting $m_\pi\to m_K$ and multiplying the resulting spectrum by the $K\to \mu \bar{\nu}_\mu$ decay branching fraction of 0.636. 

For the two-body decay of ultrarelativistic pions, the spectrum of the final product given by
\begin{equation}
\frac{dN_\ell}{dE_\ell}(E_\ell) = \int_{E_{\pi\, {\rm min}}^\ell} ^{E_{\pi\, {\rm max}}^\ell} dE_\pi \frac{dN_\pi}{dE_\pi} F_{\pi\to \ell} (E_\ell,E_\pi)\, ,
\label{DecayConvolution}
\end{equation}
where $\ell = \mu,\ \nu$. In fact, the decay spectra for the 2 body pion decays are
\begin{equation}
F_{\pi\to\mu}(E_\mu,E_\pi) = F_{\pi\to\nu}(E_\nu,E_\pi) = \frac{1}{E_\pi}\frac{1}{1-\lambda_\pi}.
\label{SinglePionDecay_Spectrum}
\end{equation}
for $\lambda_\pi=(m_\mu/m_\pi)^2$. The kinematic constraints are that $E_\mu/E_\pi>\lambda_\pi$ and $E_{\nu}/E_\pi<1-\lambda_\pi$,
so 
\begin{eqnarray}
E_{\pi\, {\rm min}}^\nu &=& E_\nu/(1-\lambda_\pi)\\
E_{\pi\, {\rm max}}^\nu &=& E_p\\
E_{\pi\, {\rm min}}^\mu &=& E_\mu\\
E_{\pi\, {\rm max}}^\mu &=& E_\mu/\lambda_\pi\ .
\end{eqnarray}

The differential neutrino spectrum from the $\pi\to\mu\to\nu$ chain in the absence of cooling is
\begin{eqnarray}
\nonumber
\frac{dN_\nu}{dE_\nu}(E_\nu) &=& \int_{E_\nu}^\infty dE_\mu \int_{E_\mu}^{E_\mu/\lambda_\pi} dE_\pi \frac{dN_\pi}{dE_\pi}\\
&\times & F_{\pi\to\mu} (E_\mu,E_\pi) F_{\mu\to\nu}(E_\nu,E_\mu)
\end{eqnarray}
where the three-body decays yield the 
distributions \cite{Barr:1989ru,Lipari93}
\begin{eqnarray}
\nonumber
F_{\mu\to\nu}(E_\nu,E_\mu) &=& \frac{1}{E_\mu}
\Biggl[ G_0
\left(\frac{E_\nu}{E_\mu}\right)
\\ 
&+& h_{\pi\to\mu}
\left(\frac{E_\mu}{E_\pi}\right)G_1\left(\frac{E_\nu}{E_\mu}
\right)
\Biggr]
\label{MuonDecaySpectrum}
\end{eqnarray}
for the neutrino spectrum from muon decay, where $h_{\pi\to \mu}$
is the $\mu^-$ polarization in $\pi^-$ decays, with
\begin{equation}
h_{\pi\to\mu}(x_\mu) = \frac{1+\lambda_\pi}{1-\lambda_\pi}-\frac{2\lambda_\pi}{(1-\lambda_\pi)x_\mu} \ ,
\label{MuonPolarization}
\end{equation}
where $x_\mu=E_\mu/E_\pi$.
Eq. \eqref{MuonDecaySpectrum} holds for $\mu^+$ decay with $h_{\pi\to\mu}\to
-h_{\pi\to\mu}$ and the identical sign change in \eqref{MuonPolarization}, so the $\nu_\mu$ distribution from $\pi^-\to \mu^- \to \nu_\mu$ is identical to the distribution of
$\bar{\nu}_\mu$ from $\pi^+\to \mu^+ \to \bar{\nu}_\mu$, and similarly for the electron neutrino and antineutrino distributions from the muon decay. The formulas for $G_0$ and $G_1$ for $\mu^-$ are summarized in Table \ref{TableA}.

\setlength{\tabcolsep}{1.0em}
\begin{table}[h]\begin{center}
\begin{tabular}{ccc}
\hline\hline
	$\mu\to\nu_\alpha$ & $G_0(y)$ & $G_1(y)$\\\hline
	$\nu_\mu$ & $\frac{5}{3}-3y^2+\frac{4}{3}y^3$ & $\frac{1}{3}-3y^2+\frac{8}{3}y^3$\\
	$\bar{\nu}_e$ & $2-6y^2+4y^3$ & $-2+12y+-18y^2+8y^3$\\
\hline\hline
\end{tabular}
\caption{Functions used in Eq. \eqref{MuonDecaySpectrum} to calculate the neutrino spectrum from $\mu^-$ decay.}
\vspace{-10pt}
\label{TableA}
\end{center}\end{table}

Here we note that the polarization of the muon is determined at production, but the neutrino spectrum is determined when the muon decays, which happens after muon cooling. 

To include cooling, we define the average polarization
\begin{equation}
\langle h_{\pi\to\mu}\rangle = \dfrac{\bigintss_{E_\mu}^{E_\mu/\lambda_\pi} dE_\pi
	\dfrac{h(E_\mu/E_\pi)}{E_\pi(1-\lambda_\pi)}\left.\dfrac{dN_\pi}{dE_\pi}\right|_\text{prod}(E_\pi)}
	{\bigintss_{E_\mu}^{E_\mu/\lambda_\pi}dE_\pi \dfrac{1}{E_\pi(1-\lambda_\pi)}\left.\dfrac{dN_\pi}{dE_\pi}\right|_\text{prod}(E_\pi)},
\end{equation}
where $dN_\pi/dE_\pi\left|_\text{prod}\right.$ is the pion spectrum at production (i.e., cooling effects are ignored). 
The neutrino spectrum is thus given by the formula
\begin{equation}
\frac{dN_\nu}{dE_\nu} = \int_{E_\nu}^\infty dE_\mu \frac{dN_\mu}{dE_\mu} \Bigg(1-\exp\Bigl(-\frac{t_{\rm cl}}{t_{\rm dec}^\mu}\Bigr)\Biggr)
\langle F_{\mu\to\nu}\rangle,
\label{PionDecay_MuSpectrum}
\end{equation}
where $dN_\mu/dE_\mu$ is found from Eq. \eqref{DecayConvolution}, with the cooling factor for the pion included in $dN/dE_\pi$ and the function $\langle F_{\mu\to\nu}\rangle$ is given by Eq.~\eqref{MuonDecaySpectrum} with the substitution $h_{\pi\to\mu}\to\langle h_{\pi\to\mu}\rangle$. This is valid under the assumption that the muons are not depolarized.


\section{Semileptonic decay formulas}
\label{app:charmdecays}
Semileptonic decay distributions as a function of neutrino energy are approximated by three-body decay formulas \cite{Pietschmann:1984en,Bhattacharya16}
with effective final state hadronic mass \cite{Bugaev98}, 
derived from the pseudoscalar three-body semileptonic decay to a lighter pseudoscalar meson, such as $D\to Kl\nu_l$. 
Neglecting lepton masses, the distribution is of the form $F_{h\to\nu}(E_{\nu},E_h) = \tilde{F}_{h\to\nu}(y)/E_h$ where $y=E_\nu/E_h$, with
\begin{eqnarray}\nonumber
\tilde{F}_{h\to\nu_l}(y)&=&\frac{1}{D(\lambda_h)}\left[6(1-2\lambda_h)(1-\lambda_h)^2-4(1-\lambda_h)^3\right.\\\nonumber
	& &-12\lambda_h^3(1-\lambda_h)+12\lambda_h^2y-6(1-2\lambda_h)y^2\\
	& & \left.+4y^3+12\lambda_h^2\ln ((1-y)/\lambda_h)\right],
\label{CharmDecayDistribution}
\end{eqnarray}
and
\begin{equation}
D(\lambda_h) = 1-8\lambda_h-12\lambda_h^2\ln\lambda_h+8\lambda_h^3-\lambda_h^4.
\end{equation}
The parameter $\lambda_h=s_h^\text{eff}/m_h^2$ is defined in terms of an effective mass $\sqrt{s_h^\text{eff}}$, shown for charm hadron decays in
Table \ref{TableA2}. The kinematic limits on $y$
are $0<y<1-\lambda_h$. 

\begin{table}[h!]\begin{center}
\begin{tabular}{cc}
\hline\hline
Decay & $\sqrt{s_h^\text{eff}}$ [GeV]\\\hline
$D^0\to\nu_l$ & 0.67\\
$D^+\to\nu_l$ & 0.63\\
$D_s^+\to\nu_l$ & 0.84\\
$\Lambda_c^+\to\nu_l$ & 1.3\\
\hline\hline
\end{tabular}
\caption{Effective masses $\sqrt{s_h^\text{eff}}$ used to calculate the neutrino spectrum from charmed hadron decay \cite{Bugaev98}.}
\label{TableA2}
\end{center}\end{table}

\vfill

\bibliography{charmingastro}

\end{document}